\documentclass[journal]{IEEEtran}

\usepackage{amsmath} 
\usepackage{amssymb}  
\usepackage{graphicx}
\usepackage{amsmath}
\usepackage{epstopdf}
\usepackage{mathtools}
\usepackage{dsfont}
\usepackage{url}
\usepackage[ruled,norelsize]{algorithm2e}
\usepackage{color}
\usepackage{booktabs} 
\usepackage{multirow}   
\usepackage{balance}    
\usepackage{soul}   
\usepackage{hyperref}   
\usepackage[noadjust]{cite} 

\makeatletter
\newcommand{\removelatexerror}{\let\@latex@error\@gobble}
\makeatother

\newcommand{\beq}{\begin{equation}}
\newcommand{\eeq}{\end{equation}}

\newcommand{\argminF}{\mathop{\mathrm{argmin}}\limits}   
\usepackage{bm}

\newcounter{algorithmctr}[section]
\renewcommand{\thealgorithmctr}{\thesection.\arabic{algorithmctr}}
   {\refstepcounter{algorithmctr}\begin{list}{}{%
       \setlength{\rightmargin}{0\linewidth}%
       \setlength{\leftmargin}{.05\linewidth}
        \setlength{\itemsep}{1pt}
  \setlength{\parskip}{0pt}
  \setlength{\parsep}{0pt}}%
       \rmfamily\small
       \item[]{\setlength{\parskip}{0ex}\hrulefill\par%
        \nopagebreak{\bfseries\textsf{Algorithm \thealgorithmctr~}}}}%
   {{\setlength{\parskip}{-1ex}\nopagebreak\par\hrulefill} \end{list}}
\IEEEoverridecommandlockouts

\title{\LARGE \bf Learning How to Autonomously Race a Car: \\a Predictive Control Approach}

\author{Ugo Rosolia and Francesco Borrelli
\thanks{U.\ Rosolia and F.\ Borrelli are with the Department of Mechanical Engineering, University of California at Berkeley ,
        Berkeley, CA 94701, USA
        {\tt\small\{ugo.rosolia, fborrelli\}{@}berkeley.edu}}%
}

\begin{document}

\maketitle
\thispagestyle{empty}
\pagestyle{empty}

\begin{abstract}
In this paper we present a Learning Model Predictive Controller (LMPC) for autonomous racing. We model the autonomous racing problem as a minimum time iterative control task, where an iteration corresponds to a lap. The system trajectory and input sequence of each lap are stored and used to systematically update the controller for the next lap. In the proposed approach the race time does not increase at each iteration. The first contribution of the paper is to propose a local LMPC which reduces the computational burden associated with existing  LMPC strategies. In particular, we show how to construct a local safe set and approximation to the value function, using a subset of the stored data. 
The second contribution is to present a system identification strategy for the autonomous racing iterative control task. We use data from previous iterations and the vehicle's kinematic equations of motion to build an affine time-varying  prediction model. The effectiveness of the proposed strategy is demonstrated by experimental results on the Berkeley Autonomous Race Car (BARC) platform. 
\end{abstract}


\section{Introduction}
Autonomous driving is an active research field. Over the past decades several techniques have been proposed for different driving scenarios \cite{rossetter2006lyapunov,MPC2,kuwata2009real,MPC3,campbell2010autonomous,gonzalez2016review,katrakazas2015real,paden2016survey,williams2016aggressive}. Depending on the control task (i.e. highway driving, urban driving, emergency maneuvers) the behavior of the vehicle can be modelled with linear or nonlinear equations of motions \cite{rajamani2011vehicle}, \cite{alleyne1997comparison}. When the nonlinearities of the vehicle are excited the control task is inevitably more challenging. In this work we are interested in designing a controller for autonomous racing which can operate the vehicle in the nonlinear regime, close to the limit of the vehicle's handling capability. We formulate the autonomous racing problem as an iterative control task, where at each iteration the controller drives the vehicle around the track trying to minimize the lap time. 

Recently several approaches have been proposed for autonomous racing. In \cite{alrifaee2018real} the authors reformulated the autonomous racing control task as a non-convex optimization problem and then proposed a linearization strategy to compute an approximate solution. The authors in \cite{verschueren2014towards} proposed a Nonlinear Model Predictive Control (NMPC) strategy which exploits a Pacejka tire model identified form experimental data. The NMPC is implemented on an experimental set-up using an  exact Hessian SQP-type optimization algorithm. NMPC strategies for autonomous racing are tested also in \cite{verschueren2016time}, where the authors compared two control methodologies based on different parametrizations of the vehicle's model.
In \cite{RacingLMS} the authors compared two approaches, the first one based on a tracking MPC and the second one based on a MPC formulated in a space dependent frame. 
A Model Predictive Contouring Control (MPCC) was presented in \cite{RacingETH}. In MPCC the controller objective is a trade-off between the progress along the track and the contouring error. First, an high level MPC computes the optimal racing trajectory. Afterward, a low level controller is used to track the optimal racing line.  This strategy is extended in \cite{liniger2017real} to design a racing controller which guarantees recursive constraint satisfaction. Also in \cite{kapania2015path} the control problem is divided in two steps. First, a reference trajectory is computed using the method proposed in \cite{theodosis2011generating}. Afterwards, an iterative learning control (ILC) approach is used for tracking. The authors showed the effectiveness of the proposed approach by experimental testing on a full size vehicle. We proposed to reformulate the autonomous racing problem as an iterative control task. The controller is not based on a precomputed racing line and it learns from experience a trajectory which minimizes the lap time. 
In particular, the closed-loop trajectories at each lap are stored and used to systematically update the controller for the next lap. 
This paper builds on~\cite{CDCRepetitiveRacing,ACCRosolia,rosolia2017learning} and has two main contributions.

The first contribution is to propose a local LMPC strategy where the terminal cost and constraint are updated at each time step. In particular at each time $t$, we exploit the planned trajectory at time $t-1$ to construct a local terminal cost and constraint. Conversely to our previous works \cite{CDCRepetitiveRacing,ACCRosolia,rosolia2017learning}, the terminal cost and constraint are computed using a subset of the stored data, therefore the proposed local LMPC enables the reduction of computational burden associated with existing LMPC strategies. The effectiveness of the proposed approach is demonstrated on the Berkeley Autonomous Race Car (BARC)\footnote{A video of the experiment can be found at \url{https://youtu.be/ZBFJWtIbtMo}} platform. We show that the proposed controller is able to improve the lap time, until it converges to a steady state behavior. Finally, we analyze the lateral acceleration acting on the closed-loop system and we confirm that the controller learns to drive the vehicle at the limit of its handling capability.

The second contribution of this work is to propose a system identification strategy tailored to the autonomous racing application. 
We propose to exploit both the kinematic equations of motion and data from previous iterations to identify an Affine Time Varying (ATV) prediction model used for control. In particular, we use a local linear regressor to learn the relationships between the inputs and the vehicle's velocities. Furthermore, we linearize the kinematic equations of motion to approximate the evolution of the vehicle's position as a function of the velocities. Conversely to our previous works \cite{CDCRepetitiveRacing,ACCRosolia},  this strategy allow us to reformulate the LMPC as a Quadratic Program (QP) which can be solved efficiently.

This paper is organized as follows: in Section II we introduce the problem formulation. Section III illustrates the LMPC design. In particular, it shows how to construct local safe sets and value function approximations using a subset of the collected data. Section IV illustrates the system identification strategy used in the experiments. Finally, in Section V we present the experimental results on the Berkeley Autonomous Race Car (BARC) platform. Section VIII provides final remarks.

\section{Problem Formulation}\label{sec:probFormulation}

\begin{figure}[h!]
	\centering
	\includegraphics[width=1.0\columnwidth]{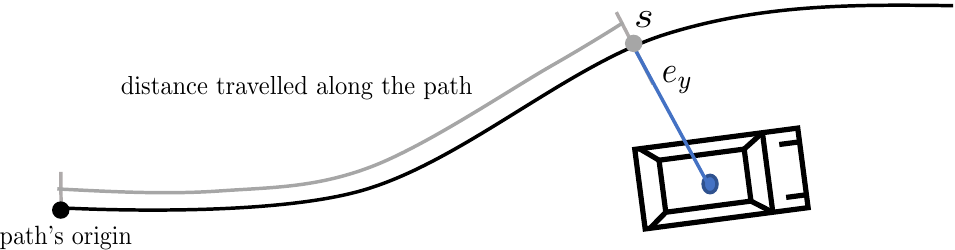}
	\caption{Representation of the vehicle's position in the curvilinear reference frame.}
	\label{Fig:CurvAbs}
\end{figure}

Consider the following  state and input vectors 
\begin{equation*}
    \begin{aligned}
        x = \begin{bmatrix} v_x , v_y, w_z, e_{\psi}, s, e_y \end{bmatrix}^\top \text{ and } u = \begin{bmatrix} \delta, a \end{bmatrix}^\top,
    \end{aligned}
\end{equation*}
where $ w_z, v_x , v_y,$ are the vehicle's yaw rate, longitudinal and lateral velocities. The position of the vehicle is represented in the curvilinear reference frame \cite{micaelli1993trajectory}, where $s$ is the distance travelled along the centerline of the track. The states $e_{\psi}$ and $e_y$ are the heading angle and lateral distance error between the vehicle and the centerline of the track, as shown in Figure~\ref{Fig:CurvAbs}. Finally, $\delta$ and $a$ are the steering and acceleration commands. The vehicle is described by the dynamic bicycle model 
\begin{equation}\label{eq:system}
x_{t+1}=f(x_t,u_t),
\end{equation}
where $f(\cdot,\cdot)$ is derived from kinematics and balancing the forces acting on the tires \cite{rajamani2011vehicle}. A detailed expression can be found in \cite[Chapter 2]{rajamani2011vehicle}. Note that in the curvilinear reference frame state and input constraints are convex, i.e.
\begin{equation*}
\begin{aligned}
&x_t \in \mathcal{X} = \{x \in \mathbb{R}^n: F_x x\leq b_x\}, \\
&u_t\in \mathcal{U}= \{u \in \mathbb{R}^d: F_u u\leq b_u\},\ \forall t \geq 0.
\end{aligned}
\end{equation*}

The goal of the controller is to drive the system from the starting point $x_S$ to the terminal set $\mathcal{X}_F$. More formally, the controller aims to solve the following minimum time optimal control problem 
\begin{equation}\label{eq:minTimeOPC}
    \begin{aligned}
        \min_{T, u_0, \ldots, u_{T-1} } & \quad \sum_{t = 0}^{T-1} 1 \\
        \text{s.t. } & \quad x_{t+1} = f(x_t, u_t), &&\forall t = [0,\dots, T-1] \\
        & \quad x_t \in \mathcal{X},\ u_t\in \mathcal{U}, &&\forall t = [0,\dots, T] \\
        & \quad x_T = \mathcal{X}_F,~ x_0 = x_S,
    \end{aligned}
\end{equation}
where for a track of length $L$ the terminal set 
\begin{equation}\label{eq:pointBeyondFinischLine}
    \mathcal{X}_F = \{x \in \mathbb{R}^n : [0~0~0~0~1~0] x = s \geq L \}
\end{equation}
represents the states beyond the finish line.


\section{Controller Design}
In this section, we first show how to use historical data to construct a terminal constraint set and terminal cost function. Afterwards, we exploit these quantities to design the controller.

\subsection{Stored Data}
As stated in the introduction, we define one iteration as a successful lap around the race track and we store the closed-loop trajectories. In particular, at the $j$th iteration we define the vectors
\begin{equation}\label{eq:storedTrajectories}
\begin{aligned}
 {\bf{u}}^j ~ &= ~ [u_0^j,\ldots,~u_{ T^j}^j] \\
 {\bf{x}}^j ~ &= ~ [x_0^j,\ldots,~x_{T^j}^j], 
\end{aligned}
\end{equation}
which collect the evolution of closed-loop system and associated input sequence. In the above definitions, $T^j$ denotes the time at which the closed-loop system reached the terminal set, i.e. $x_{T^j} \in \mathcal{X}_F$.

\subsection{Local Convex Safe Set}\label{sec:CVXlocalSS}
In this section, we define the local convex safe set. Differently from our previous works~\cite{rosolia2017learning,CDCRepetitiveRacing,ACCRosolia}, this quantity is constructed using a subset of the stored data points. In particular, the local convex safe set around $x$ is defined as the convex hull of the $K$-nearest neighbors to $x$.

First, for the $j$th trajectory we define the set of time indices $[t^{j,*}_1, \ldots, t^{j,*}_K]$ associated with the $K$-nearest neighbors to the point $x$,
\begin{equation}\label{eq:timeInLS}
\begin{aligned}
    [t^{j,*}_1, \ldots, t^{j,*}_K] = \argminF_{t_1, \ldots, t_K}   & \quad \sum_{i = 1}^K ||x^j_{t_i} - x||_D^2 \\
    \text{s.t.}  & \quad t_i \neq t_k,~ \forall i \neq k \\
    & \quad t_i \in \{0, \ldots, T^j \}, \forall i \in \{1, \ldots,K \}.
\end{aligned}
\end{equation}
In the above definition $||y||_D^2 = y^\top D^\top D y $ for the user-defined matrix $D$, which may be chosen to take into account the relative scaling or relevance of different variables. We chose $D= diag(0,0,0,0,1,0)$ to select the $K$-nearest neighbors with respect to the curvilinear abscissa $s$, which represents a proxy for the distance between two stored data points of the same lap. Furthermore, as the vehicle moves forward on the track, at each lap the stored data are ordered with respect to the travelled distance $s$ and the computation of \eqref{eq:timeInLS} is simplified. The $K$-nearest neighbors to $x$ from the $l$th to the $j$th iteration are collected in the following matrix
\begin{equation*}
    D_l^j(x) = [x_{t^{l,*}_1}^l, \ldots, x_{t^{l,*}_K}^l, \ldots, x_{t^{j,*}_1}^j, \ldots, x_{t^{j,*}_K}^j],
\end{equation*}
which is used to define the local convex safe set around $x$
\begin{equation}\label{eq:LS}
\begin{aligned}
    \mathcal{CL}^j_l(x) = \{  \bar x \in  \mathbb{R}^n : \exists {\boldsymbol{\lambda}} \in & \mathbb{R}^{K(j-l+1)}, 
    \\ & \boldsymbol{\lambda} \geq 0, \mathds{1}{\boldsymbol{\lambda}} = 1,  D_l^j(x) \boldsymbol{\lambda} = \bar x\}.
\end{aligned}
\end{equation}
Notice that the above local convex safe set $\mathcal{CL}^j_l(x)$ represents the convex hull of the $K$-nearest neighbors to $x$ from the $l$th to $j$th iteration.

Finally, we define the matrix 
\begin{equation*}
    S_l^j(x) = [x_{t^{l,*}_1+1}^l, \ldots, x_{t^{l,*}_K+1}^l, \ldots, x_{t^{j,*}_1+1}^j, \ldots, x_{t^{j,*}_K+1}^j] 
\end{equation*}
which collects the evolution of the states stored in the columns of the matrix $D_l^j(x)$. The above matrix $S_l^j(x)$ will be used in Section~\ref{sec:LMPC} to construct the local convex safe set at each time step. 

\subsection{Local Convex Q-function}
In this section, we exploit the stored data to construct an approximation to the cost-to-go over the local convex safe set $\mathcal{CL}^j_l(x)$ around $x$. In particular, we define the local convex $Q$-function around $x$ as the convex combination of the cost associated with the stored trajectories,
\begin{equation}\label{eq:localQfun}
    \begin{aligned}
         Q_l^j(\bar x, x) = \min_{\boldsymbol{\lambda} } & \quad {\bf{J}}^j_l(x) \boldsymbol{\lambda} \\
        \text{s.t }& \quad \boldsymbol{\lambda} \geq 0, ~ \mathds{1}{\boldsymbol{\lambda}} = 1,  D_l^j(x) \boldsymbol{\lambda} = \bar x,
    \end{aligned}
\end{equation}
where $\boldsymbol{\lambda} \in \mathbb{R}^{k(j-l)}$, $\mathds{1}$ is a row vector of ones and the row vector
\begin{equation*}
\begin{aligned}
    {\bf{J}}^j_l(x) = [J_{t^{l,*}_1 \rightarrow T^l}^l &(x_{t^{l,*}_1}^l), \ldots, J_{t^{l,*}_M\rightarrow T^l}^l(x_{t^{l,*}_M}^l), \ldots, \\&J_{t^{j,*}_1\rightarrow T^j}^j(x_{t^{j,*}_1}^j), \ldots, J_{t^{j,*}_M\rightarrow T^j}^j(x_{t^{j,*}_M}^j)],
\end{aligned}
\end{equation*}
collects the cost-to-go associated with the $K$-nearest neighbors to $x$ from the $l$th the $j$th iteration. The cost-to-go $J_{t \rightarrow T^j}^j (x_{t}^j) = T^j-t$ represents the time to drive the vehicle from $x_t^j$ to the finish line along the $j$th trajectory. We underline that the cost-to-go is computed after completion of the $j$th iteration.

\subsection{Local LMPC Design}\label{sec:LMPC}
The local convex safe set and the local convex $Q$-function are used to design the controller. At each time $t$ of the $j$th iteration the controller solves the following finite time optimal control problem
\begin{subequations}\label{eq:FTOCP}
	\begin{align}
    J_{t\rightarrow t+N}^{\scalebox{0.4}{LMPC},j}(&x_t^j, z_t^j) = \notag \\
    \min_{{\bf{U}}_t^j, \boldsymbol{\lambda}_t^j}  \quad &\bigg[  \sum_{k=t}^{t+N-1}  h(x_{k|t}^j) +{\bf{J}}^{j-1}_l(z_t^j) \boldsymbol{\lambda}_t^j\bigg] \label{eq:FTOCP_Const}\\
	\text{s.t.}\quad &x_{t|t}^j=x_t^j, \label{eq:FTOCP_IC}\\
	&\boldsymbol{\lambda}_t^j \geq 0, \mathds{1}{\boldsymbol{\lambda}}_t^j = 1,  D_l^{j-1}(z_t^j) \boldsymbol{\lambda}_t^j = x_{t+N|t}^j \label{eq:FTOCP_Term}\\
    &x_{k+1|t}^j=A_{k|t}^j x_{k|t}^j + B_{k|t}^j u_{k|t}^j + C_{k|t}^j, \label{eq:FTOCP_Dyn} \\
	&x_{k|t}^j \in \mathcal{X}, u_{k|t}^j \in \mathcal{U}, \label{eq:FTOCP_Cons}\\ 
	&\forall k = t, \cdots, t+N-1, \notag
	\end{align}
\end{subequations}
where ${\bf{U}}_t^j = [u_{t|t}^j,\ldots,u_{t+N-1|t}^j] \in \mathbb{R}^{d \times N}$, $\boldsymbol{\lambda}_t^j \in \mathbb{R}^{(j-l+1)K}$ and the stage cost in~\eqref{eq:FTOCP_Const} 
\begin{equation*}
    h(x) = \begin{cases} 1 & \mbox{If } x \notin \mathcal{X}_F\\
    0 & \mbox{Else }\\
    \end{cases}.
\end{equation*}
In the above finite time optimal control problem equations \eqref{eq:FTOCP_IC},  \eqref{eq:FTOCP_Dyn} and  \eqref{eq:FTOCP_Cons} represent the dynamic update, state and input constraints. Finally, \eqref{eq:FTOCP_Term} enforces $x_{t+N|t}^j$ into the local convex safe set defined in Section~\ref{sec:CVXlocalSS}. 
The optimal solution to \eqref{eq:FTOCP} at time $t$ of the $j$th iteration
\begin{equation}\label{eq:optSolutionLMPC}
\begin{aligned}
    \boldsymbol{\lambda}_t^{j,*}, [x_{t|t}^{j,*},\ldots,x_{t+N|t}^{j,*}] \text{ and } {\bf{U}}_t^{j,*} &= [u_{t|t}^{j,*},\ldots,u_{t+N-1|t}^{j,*}]
\end{aligned}
\end{equation}
is used to compute the following vector
\begin{equation}\label{eq:z_t^j}
    z_{t}^j = \begin{cases} x_{N}^{j-1} & \mbox{If } t = 0\\
    S_l^j(z_{t-1}^j) \boldsymbol{\lambda}_{t-1}^{j,*} & \mbox{Otherwise }\\
    \end{cases},
\end{equation}
which at time $t$ defines the local convex safe set $\mathcal{LS}^j_l(z_t^j)$ and local $Q$-function $Q^j_l(x, z_t^j)$ in \eqref{eq:FTOCP}. The above vector $z_t^j$ represents a candidate terminal state for the planned trajectory of the LMPC at time $t$. First, we initialize the candidate terminal state $z_0^j$ using the $(j-1)$th trajectory. Afterwards, we update the vector $z_t^j$ as the convex combination of the columns of the matrix $S_l^j(z_t^j)$ from Section~\ref{sec:CVXlocalSS}. Notice that if the systems is linear or if a linearized system approximates the nonlinear dynamics over the local convex safe set, then there exists a feasible input which drives the system from $x_{t+N|t}^{j,*} = D_{t}^{j-1}(z_t^j)\boldsymbol{\lambda}_{t}^{j,*}$ to $z_{t+1}^j = S_l^{j-1}(z_{t}^j) \boldsymbol{\lambda}_{t}^{j,*}$.

Finally, we apply to the system \eqref{eq:system} the first element of the optimizer vector,
\begin{equation}\label{eq:policyLMPC}
    u_t^j = u_{t|t}^{j,*}.
\end{equation}
The finite time optimal control problem \eqref{eq:FTOCP} is repeated at time $t+1$, based on the new state $x_{t+1|t+1} = x_{t+1}^j$.
\section{System Identification Strategy}

In this section, we illustrate the system identification strategy used to build an Affine Time Varying (ATV) model which approximates the vehicle dynamics. First, we introduce the kinematic equations of motion which describe the evolution of the vehicle's position as a function of the velocities. Afterwards, we present the strategy used to approximate the dynamic equations of motion, which model the evolution of the vehicle's velocities as a function of the input commands. Finally, we describe the ATV model, which is computed online linearizing the kinematic equations of motion and evaluating the approximate dynamic equations of motion along the shifted optimal solution to the LMPC.

\subsection{Kinematic Model}
As mentioned in Section \ref{sec:probFormulation}, the position of the vehicle is expressed in the Frenet reference frame \cite{micaelli1993trajectory}. In particular, we describe the position of the vehicle in terms of lateral distance $e_y$ from the centerline of the road and distance $s$ traveled along a predefined path (Fig. \ref{Fig:CurvAbs}). The state $e_{\psi}$ represents the difference between the vehicle's heading angle and the angle of the tangent vector to the path at the curvilinear abscissa $s$. 

The rate of change of the vehicle's position in the curvilinar reference frame is described by the following kinematic relationships 
\begin{equation*}
	\begin{aligned}
	\dot{e}_{\psi} &= w_z - \frac{v_x \cos(e_{\psi})  - v_y \sin(e_{\psi})}{ 1 - \kappa(s) e_y}\kappa(s)  \\
	\dot s &= \frac{v_x \cos(e_{\psi})  - v_y \sin(e_{\psi})}{1 -  \kappa(s) e_y}\\
	\dot{e}_y &= v_x \sin(e_{\psi})  + v_y \cos(e_{\psi}),\\
	\end{aligned}
\end{equation*}
where $\kappa(s)$ is the curvature of the centerline of the track at the curvilinear abscissa $s$ \cite{micaelli1993trajectory}. The above equations can be Euler discretized to approximate the vehicle's motion as a function of the vehicle's velocities
\begin{equation}\label{eq:kinematicEquationMotion}
	\begin{aligned}
	{e}_{\psi_{k+1}} &= f_{e_\psi}(x_k) = e_{\psi_k} \\
	&\quad+ dt \Bigg( w_{z_k} - \frac{v_{x_k} \cos(e_{\psi_k})  - v_{y_k} \sin(e_{\psi_k})}{1 - \kappa(s_k) e_{y_k}} \kappa(s_k)\Bigg) \\
	s_{k+1} &=f_s(x_k) = s_k +  dt \Bigg(\frac{v_{x_k} \cos(e_{\psi_k})  - v_{y_k} \sin(e_{\psi_k})}{1 - \kappa(s_k) e_{y_k}} \Bigg)\\
	\dot{e}_y &= f_{e_y}(x_k) = e_{y_k} + dt \Bigg(v_{x_k} \sin(e_{\psi_k})  + v_{y_k} \cos(e_{\psi_k}) \Bigg),\\
	\end{aligned}
\end{equation}
where $dt$ is the discretization time. The above equations will be linearized to compute an ATV prediction model. It is interesting to notice that equations \eqref{eq:kinematicEquationMotion} are independent of the vehicle's physical parameters, because these are derived from kinematic relationships between velocities and position. 

\subsection{Dynamic Model}
The dynamic equations of motion, which describe the evolution of the vehicle's velocities, may be computed balancing the forces acting on the tires \cite{rajamani2011vehicle}. Therefore, the dynamic equations depend on physical parameters associated with the vehicle, tires and asphalt. These parameters may be estimated through a system identification campaign. However, the nonlinear dynamic equations of motion should be linearized in order to obtain an ATV model which allows us to reformulate the  LMPC as a QP. Instead of identifying the parameters of a nonlinear model and then linearize it, we propose to directly learn a linear model around $x$ using a local linear regressor.  We introduce the Epanechnikov kernel function \cite{epanechnikov1969non}
\begin{equation*}
	K(u) = \begin{cases}
	\frac{3}{4}(1-u^2), &\mbox{ for } |u| < 1 \\
	0, &\mbox{ else }
	\end{cases},
\end{equation*}
which is used to compute a local linear model around $x$ for the longitudinal and lateral dynamics. In particular, for $l = \{v_x, v_y, w_z\}$ we compute the following regressor vector
\begin{equation}\label{eq:locaLinRef}
	\begin{aligned}
	\Gamma^{l}(x) = \argminF_{\Gamma} \sum_{\{k,j\} \in I(x)} K\Bigg( \frac{||x- x_k^j||_Q^2}{h} \Bigg) y_k^{j,l}(\Gamma),
	\end{aligned}
\end{equation}
where the hyperparameter $h \in \mathbb{R}_+$ is the bandwidth, the row vector $\Gamma \in \mathbb{R}^5$,
\begin{equation*}
	\begin{aligned}
	y_k^{j, v_x}(\Gamma) &= ||v_{x_{k+1}}^j - \Gamma [
	v_{x_k}^j,~ 
	v_{y_k}^j,~ 
	w_{z_k}^j,~ 
	a_k^j,~1]^T|| \\
	y_k^{j, v_y}(\Gamma) &= ||v_{y_{k+1}}^j - \Gamma [
	v_{x_k}^j,~ 
	v_{y_k}^j,~ 
	w_{z_k}^j,~ 
	\delta_k^j,~1]^T|| \\
	y_k^{j, w_z}(\Gamma) &= ||w_{z_{k+1}}^j - \Gamma [
	v_{x_k}^j,~ 
	v_{y_k}^j,~ 
	w_{z_k}^j,~ 
	\delta_k^j,~1]^T||,
	\end{aligned}
\end{equation*}
and $I^j_l(x)$ is the set of indices
\begin{equation*}
	\begin{aligned}\label{eq:SysID_Idef}
	I^j_l(x) = \argminF_{\{k_1, j_1\}, \ldots, \{k_P, j_P\} } & \quad \sum_{i = 1}^{P} ||x - x_{k_i}^{j_i}||_Q^2\\
	\text{s.t.} & \\
    & \quad k_i \neq k_n,~ \forall j_i = j_n \\
    & \quad k_i \in \{1, 2, \ldots \}, \forall i \in \{1, \ldots,P \} \\
    & \quad j_i \in \{l, \ldots, j \}, \forall i \in \{1, \ldots,P \},
	\end{aligned}
\end{equation*}
where $||y||_Q = y^\top Q^\top Q y$ and the matrix $Q$ is user defined.
For the stored data from iteration $l$ to iteration $j$, the set $I^j_l(x)$ collects the indices associated with the $P$-nearest neighbors to the state $x$. Finally, the user-defined matrix $Q$ takes into account the relative scaling of different variables.

Notice that the optimizer in \eqref{eq:locaLinRef} can be used to approximate the evolution of vehicle's velocities,
\begin{equation}\label{eq:dynEquationsMotion}
    \begin{aligned}
        \begin{bmatrix}
        v_{x_{k+1}} \\
        v_{y_{k+1}} \\
        w_{z_{k+1}}
        \end{bmatrix}& = \begin{bmatrix}
        \Gamma_{1:3}^{v_x}(x) \\
        \Gamma_{1:3}^{v_y}(x) \\
        \Gamma_{1:3}^{w_z}(x)
        \end{bmatrix} 
        \begin{bmatrix}
        v_{x_k} \\ v_{y_k} \\ w_{z_k}
        \end{bmatrix}
         \\& \quad + \begin{bmatrix}
        \Gamma_{4}^{v_x}(x) & 0 \\
        0 & \Gamma_{4}^{v_y}(x) \\
        0 &  \Gamma_{4}^{w_z}(x)
        \end{bmatrix}
        \begin{bmatrix}
        a_k \\ \delta_k
        \end{bmatrix}
        +\begin{bmatrix}
        \Gamma_{5}^{v_x}(x) \\
        \Gamma_{5}^{v_y}(x) \\
        \Gamma_{5}^{w_z}(x)
        \end{bmatrix},
    \end{aligned}
\end{equation}
where for $l = \{v_x, v_y, w_z\}$ the scalar $\Gamma_{i}^{l}(x)$ denotes the $i$th element of the vector $\Gamma^{l}(x)$ and $\Gamma_{1:3}^{l}(x) \in \mathbb{R}^3$ is a row vector collecting the first three elements of $\Gamma^{l}(x)$ in~\eqref{eq:locaLinRef}.

\subsection{Affine Time Varying Model}
In this section we describe the strategy used to build an ATV model, which is then used for control. At time $t$ of the $j$th iteration we define the candidate solution $\bar {\bf{x}}_{t}^j = [\bar x_{t|t}^{j}, \ldots, \bar x_{t+N|t}^{j}]$ to Problem~\eqref{eq:FTOCP} using the optimal solution at time $t-1$ from~\eqref{eq:optSolutionLMPC},
\begin{equation*}
    \bar x_{k|t}^{j} = \begin{cases} 
    x_{k|t-1}^{j,*} & \mbox{If } k \in \{t, \ldots, t + N - 1\} \\
    z_{t}^{j} & \mbox{If } k = t+N
    \end{cases}.
\end{equation*}
Finally at each time $t$ of iteration $j$, the above candidate solution is used to build the following ATV model
\begin{equation}\label{eq:SysID_Model}
	x_{k+1|t}^j = A_{k|t}^j x_{k|t}^j + B_{k|t}^j u_{k|t}^j + C_{k|t}^j,
\end{equation}
where $x_{k|t}^j = [v_{x_{k|t}}^j, v_{y_{k|t}}^j, w_{y_{k|t}}^j, e_{\psi_{k|t}}^j, s_{{k|t}}^j, e_{y_{k|t}}^j]$ and the matrices $A_{k|t}^j$, $B_{k|t}^j$ and $C_{k|t}^j$ are obtained linearizing \eqref{eq:kinematicEquationMotion} around $\bar x_{k|t}^{j}$ and evaluating \eqref{eq:dynEquationsMotion} at $\bar x_{k|t}^{j}$, 
\begin{equation}\label{eq:matricesLoc1}
	\begin{aligned}
	A_{k|t}^j = \begin{bmatrix}
	\Gamma^{v_x}_{1:3}(\bar x_{k|t}^{j}) ~~ 0 ~~ 0 ~~ 0 \\
	\Gamma^{v_y}_{1:3}(\bar x_{k|t}^{j}) ~~ 0 ~~ 0 ~~ 0 \\
	\Gamma^{w_z}_{1:3}(\bar x_{k|t}^{j}) ~~ 0 ~~ 0 ~~ 0 \\
	~~(\nabla_x f_{{e}_\psi}(x)|_{\bar x_{k|t}^{j}})^\top\\
	~~(\nabla_x f_{s}(x)|_{\bar x_{k|t}^{j}})^\top\\
	~~(\nabla_x f_{e_y}(x)|_{\bar x_{k|t}^{j}})^\top\\
	\end{bmatrix}, B_{k|t}^j = 
	\begin{bmatrix}
	\Gamma^{v_x}_4(\bar x_{k|t}^{j}) ~~~\, 0 ~~~\\
	\,0  \,~~~~~~~ \Gamma^{v_y}_4(\bar x_{k|t}^{j}) \\
    \,0  \,\,~~~~~~ \Gamma^{w_z}_4(\bar x_{k|t}^{j}) \\
	0  ~~~~~~~~~~~~~ 0 ~~~\\
	0  ~~~~~~~~~~~~~ 0 ~~~\\
	0  ~~~~~~~~~~~~~ 0 ~~~\\
		\end{bmatrix}
	\end{aligned}
\end{equation}
and 
\begin{equation}\label{eq:matricesLoc2}
	\begin{aligned}
        C_k &= \begin{bmatrix}
		\Gamma^{v_x}_5(\bar x_{k|t}^{j}) \\
		\Gamma^{v_y}_5(\bar x_{k|t}^{j}) \\
		\Gamma^{w_z}_5(\bar x_{k|t}^{j}) \\
		f_{e_{y}}(\bar x_{k|t}^{j}) - (\nabla_x f_{e_{y}}(x)|_{\bar x_{k|t}^{j}})^\top \bar x_{k|t}^{j} \\	
		f_s (\bar x_{k|t}^{j}) - (\nabla_x f_s (x)|_{\bar x_{k|t}^{j}})^\top \bar x_{k|t}^{j} \\
		f_{e_{\psi}} (\bar x_{k|t}^{j}) - (\nabla_x f_{e_{\psi}} (x)|_{\bar x_{k|t}^{j}})^\top \bar x_{k|t}^{j}
		\end{bmatrix}
	\end{aligned}.
\end{equation}


\section{Results}
The proposed control strategy has been implemented on a 1/10-scale open source vehicle platform called Berkeley Autonomous Race Car\footnote{A video of the experiment can be found at \url{https://youtu.be/ZBFJWtIbtMo}} (BARC). The vehicle is equipped with a set of sensors, actuators and two on-board CPUs to perform low-level control of the actuators as well as communication with a laptop, on which the high-level control is implemented. 
The CPUs are an Arduino Nano for low-level control of the actuators and an Odroid XU4 for WiFi communication with the i7 MSI GT72 laptop. 
The actuators are an electrical motor and a servo for the steering. The control architecture has been implemented in the Robot Operating System (ROS) framework, using Python and OSQP \cite{stellato2018osqp}. The code is available online\footnote{The code is available on the BARC GitHub repository in the ``devel-ugo" branch \href{https://github.com/MPC-Berkeley/barc/tree/devel-ugo}{(github.com/MPC-Berkeley/barc)}}. 

\begin{figure}[h!]
\centering
\includegraphics[width=\columnwidth]{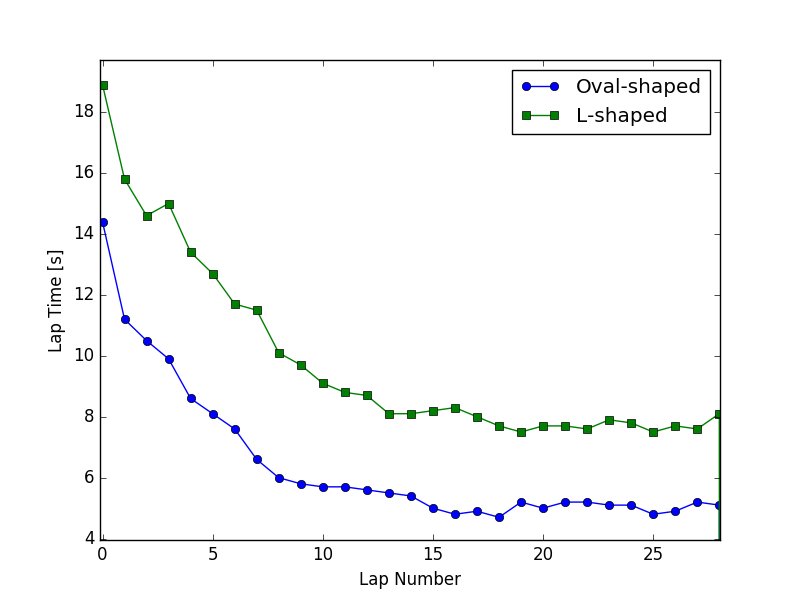}
\caption{Lap time of the LMPC on the oval-shaped and L-shaped tracks.}
\label{fig:lapTime}
\end{figure}

We initialize the algorithm performing two laps of path following at constant speed. Each $j$th iteration collects the data of two consecutive laps. Therefore, the local safe set and local $Q$-function are defined also beyond the finish line. This strategy allows us to implement the LMPC for the repetitive autonomous racing control task, as shown in \cite{CDCRepetitiveRacing}. At each $j$th lap, we use the LMPC \eqref{eq:FTOCP} and \eqref{eq:policyLMPC} to drive the vehicle from the starting line to the finish line and we use the closed-loop data to update the controller for the next lap. The parameters which define the controller are reported in Table~\ref{table:parameters}. We also added a small input rate cost in order to guarantee a unique solution to the QP associated with the LMPC.


\begin{table}[h!]
\centering\caption{Parameters used in the controller design.}\label{table:parameters}
\begin{tabular}{l|l}
 $l$  & $j-2$  \\
 $K$ & 20  \\
 $T$ & diag$(0,0,0,0,1,0)$  \\
 $Q$ & diag$(0.1,1,1,0,0,0)$ \\
 $P$ & 80 \\
 $h$ & 10 \\
 $N$ & 12 \\
\end{tabular}
\end{table}

\begin{figure*}[h!]
	\includegraphics[width=1.0\textwidth]{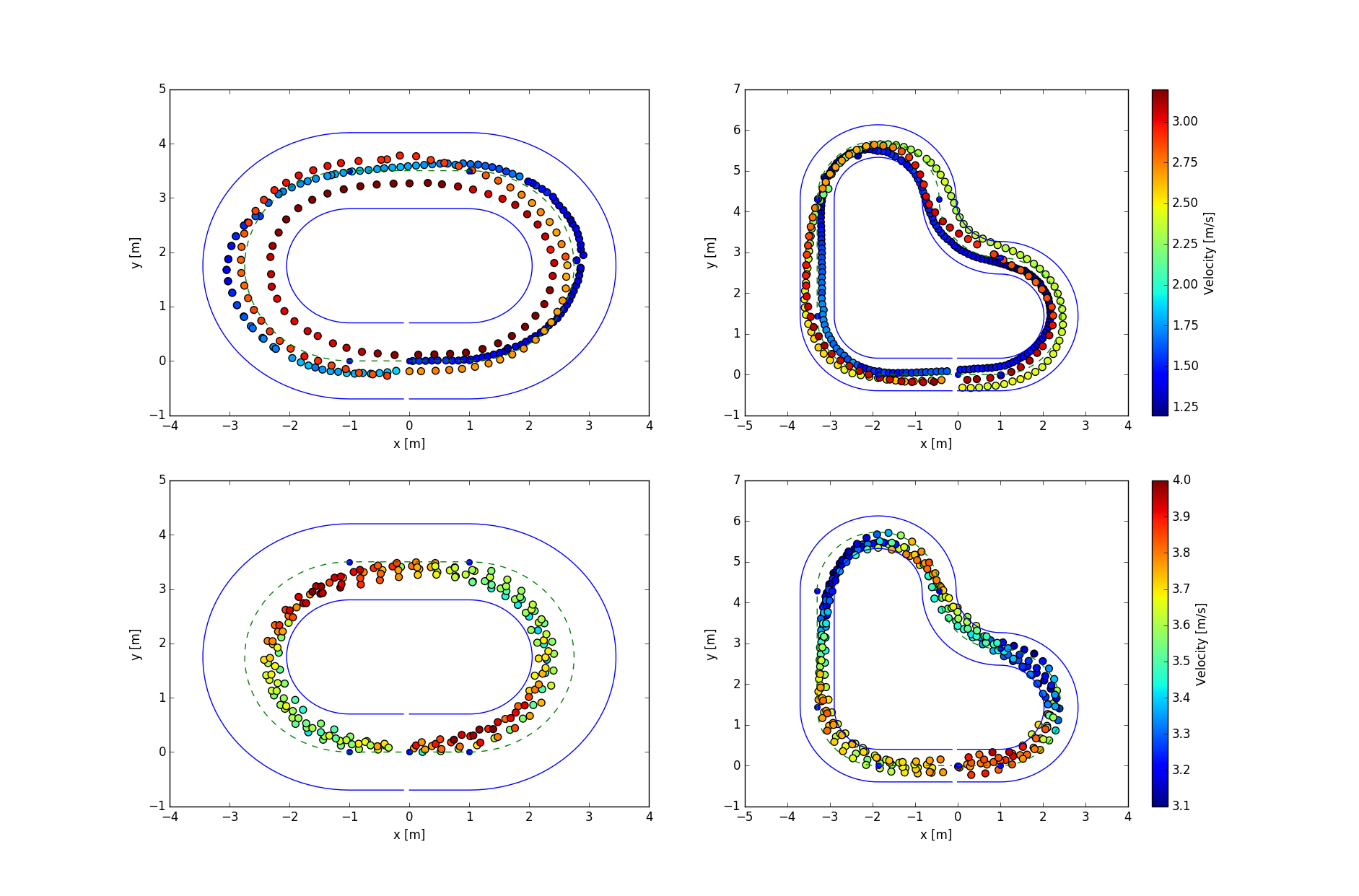}
	\caption{The first row in the above figure shows the closed-loop trajectory used to initialize the LMPC and the closed-loop trajectories after few laps of learning. The second row shows the steady state trajectories at which the LMPC has converged. Notice that the scale of the color bar changes from the first to the second row, as the vehicle runs at higher speed after the learning process has converged.}
	\label{fig:TrajectoryEvolution}
\end{figure*}

We tested the controller on an oval-shaped and L-shaped tracks on which the vehicle runs in the counter-clockwise direction. Figure~\ref{fig:lapTime} shows that the lap time decreases until convergence is reached after $29$ laps. Furthermore, Figure~\ref{fig:TrajectoryEvolution} shows the evolution of the closed-loop trajectory on the X-Y plane and the velocity profile which is color coded. In the first row we reported the path following trajectory used to initialize the LMPC and the closed-loop trajectories at laps $7$ and $15$. We notice that the controller deviates from the initial feasible trajectory (reported in blue as the vehicle speed is $1.2$m/s) in order to explore the state space and to drive the vehicle at higher speeds, until it converges to a steady-state behavior. The steady-state trajectories from lap $30$ to $34$ are reported in the bottom row of Figure~\ref{fig:TrajectoryEvolution}. Notice that the color bar representing the velocity profile changed from the first to second row as the vehicle runs at higher speed at the end of the learning process. We underline that the controller understands the benefit of breaking right before entering the curve and of accelerating when exiting. 
This behavior is optimal in racing as shown in~\cite{Stanford}.

\begin{figure}[h!]
\centering
\includegraphics[width=1.0\columnwidth]{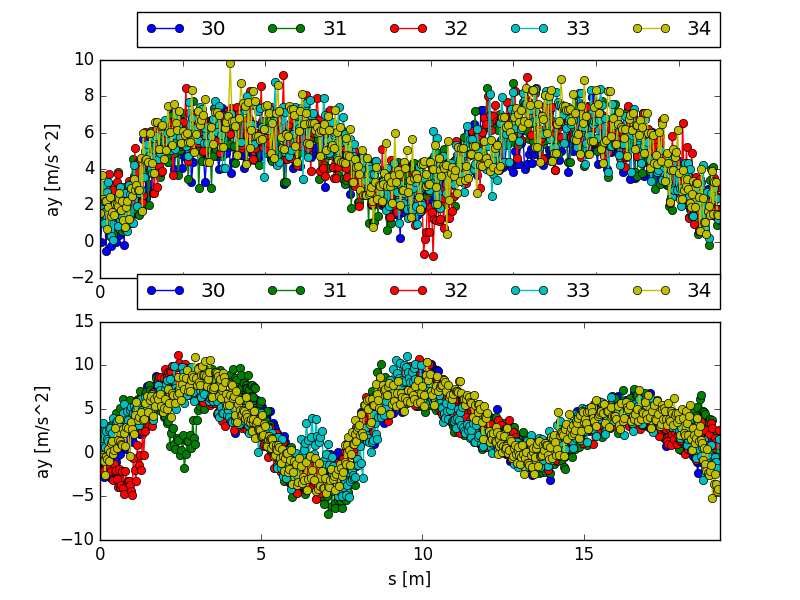}
\caption{Recorded lateral acceleration of the vehicle running on the oval-shaped track (top row) and L-shaped track (bottom row).}
\label{fig:acceleration}
\end{figure}

Figure~\ref{fig:acceleration} shows the raw acceleration measurements from the IMU. We confirm that controller is able to operate the vehicle at the limit of its handling capability, reaching a maximum lateral acceleration close to $1$g \footnote{The maximum allowed lateral acceleration is computed assuming that the aerodynamic effects are negligible and the that lateral force acting on the vehicle is $F = \mu m g$ for the friction coefficient $\mu =1$.}. 

\begin{figure}[h!]
\centering
\includegraphics[width=1.0\columnwidth]{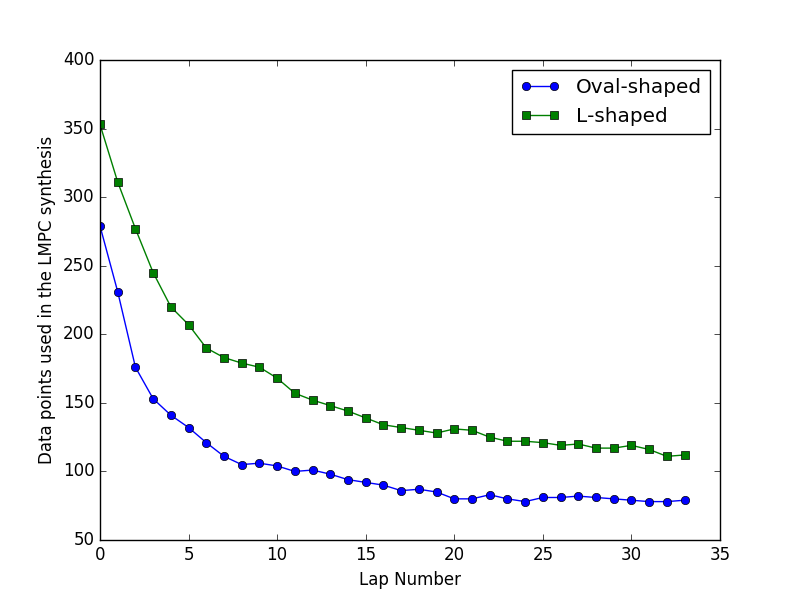}
\caption{Data points used in the LMPC design at each lap.}
\label{fig:dataUsed}
\end{figure}

\begin{figure}[h!]
\centering
\includegraphics[width=1.0\columnwidth]{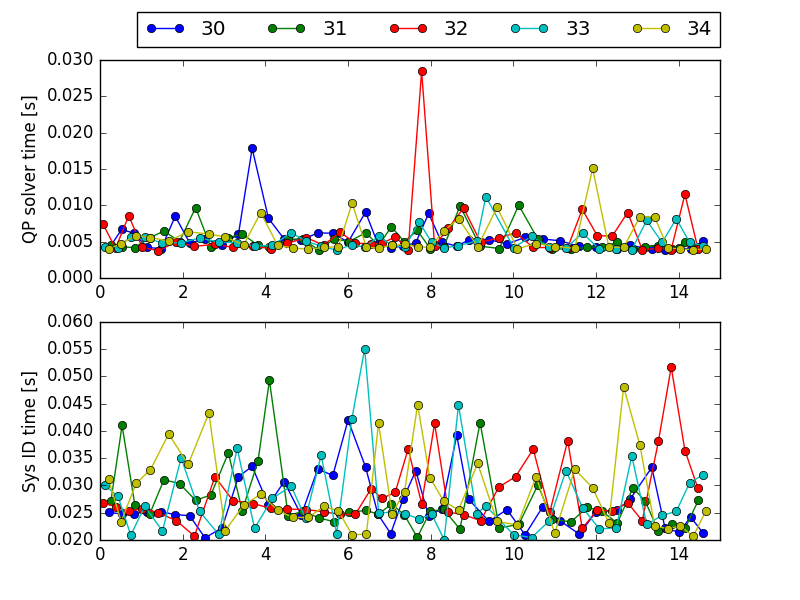}
\caption{The first rows shows the computational cost associated with the FTOCP. In the second row we reported the computational cost associated with the system identification strategy.}
\label{fig:computationaltime}
\end{figure}

Furthermore, Figure~\ref{fig:dataUsed} shows the data points used to design the LMPC. Recall from Table~\ref{table:parameters} that at the $j$th lap the LMPC policy is synthesized using the trajectories from lap $l=j-2$ to lap $j-1$. Therefore, as the controller drives faster on the track, less data points are needed to design the LMPC.
Moreover, in Figure~\ref{fig:computationaltime} we reported the computational time. It is interesting to notice that on average the finite time optimal control problem \eqref{eq:FTOCP} is solved in less then $10$ms, whereas it took $90$ms to solve the finite time optimal control problem associated with \cite{CDCRepetitiveRacing}. We underline that both strategies have been tested with a prediction horizon of $N=12$ and a sampling time of $10$Hz. This shows the advantage of using the local convex safe set in \eqref{eq:LS}, instead of the polynomial approximation to the safe set used in \cite{ACCRosolia,CDCRepetitiveRacing}. For more details on the polynomial approximation to the safe set we refer to \cite{ACCRosolia}. Finally, we notice that it would be possible to parallelize the computation of the $N-1$ linear models which define the ATV model from \eqref{eq:SysID_Model}. Indeed, at time $t$ Equations~\eqref{eq:matricesLoc1}-\eqref{eq:matricesLoc2} may be evaluated independently and in parallel for each predicted time $k$.



\section{Conclusions}\label{sec:conclusions}
We presented a Learning Model Predictive Controller (LMPC) for autonomous racing. The proposed control framework uses historical data to construct safe sets and approximations to the value function. These quantities are systematically updated when a lap is completed, as a result the LMPC learns from experience to safely drive the vehicle at the limit of handling. We demonstrated the effectiveness of the proposed strategy on the Berkeley Autonomous Race Car (BARC) platform.  Experimental results show that the controller learns to drive the vehicle aggressively, in order to minimize the lap time. In particular, the closed-loop system converged to a steady-state trajectory which cuts curves and reaches a lateral acceleration close to $1$g.


\renewcommand{\baselinestretch}{0.995}
\bibliographystyle{IEEEtran}
\bibliography{IEEEabrv,mybibfile}

\end{document}